\documentclass[conference]{IEEEtran}
\IEEEoverridecommandlockouts

\usepackage{cite}
\usepackage{amsmath,amssymb,amsfonts}
\usepackage{algorithmic}
\usepackage{graphicx}
\usepackage{multirow}
\usepackage{threeparttable}
\usepackage{float}
\usepackage{textcomp}
\usepackage{CJKutf8}
\usepackage{booktabs}
\usepackage{xcolor}
\usepackage{titlesec}
\usepackage{authblk}
\def\BibTeX{{\rm B\kern-.05em{\sc i\kern-.025em b}\kern-.08em
    T\kern-.1667em\lower.7ex\hbox{E}\kern-.125emX}}
\begin{document}
\title{\fontsize{20pt}{10pt}\selectfont SelectiveFinetuning: Enhancing Transfer Learning in Sleep Staging through Selective Domain Alignment
}
\author{Siyuan Zhao\textsuperscript{1}, Chenyu Liu\textsuperscript{2,*}, Yi Ding\textsuperscript{2,*}, Xinliang Zhou\textsuperscript{2,*}}

\affil{\textsuperscript{1}\textit{School of Computer and Information Technology, Beijing Jiaotong University, Beijing, China}}
\affil{\textsuperscript{2}\textit{College of Computing and Data Science, Nanyang Technological University, Singapore}}
\affil{Emails: siyuan012@bjtu.edu.cn; \{chenyu003, xinliang001\}@e.ntu.edu.sg; \ ding.yi@ntu.edu.sg }

\renewcommand\Authands{, } 

\maketitle

\let\thefootnote\relax\footnotetext{* Corresponding Authors}
\begin{abstract}
In practical sleep stage classification, a key challenge is the variability of EEG data across different subjects and environments. Differences in physiology, age, health status, and recording conditions can lead to domain shifts between data. These domain shifts often result in decreased model accuracy and reliability, particularly when the model is applied to new data with characteristics different from those it was originally trained on, which is a typical manifestation of negative transfer. To address this, we propose SelectiveFinetuning in this paper. Our method utilizes a pre-trained Multi-Resolution Convolutional Neural Network (MRCNN) to extract EEG features, capturing the distinctive characteristics of different sleep stages. To mitigate the effect of domain shifts, we introduce a domain aligning mechanism that employs Earth Mover’s Distance (EMD) to evaluate and select source domain data closely matching the target domain. By finetuning the model with selective source data, our SelectiveFinetuning enhances the model's performance on target domain that exhibits domain shifts compared to the data used for training. Experimental results show that our method outperforms existing baselines, offering greater robustness and adaptability in practical scenarios where data distributions are often unpredictable.
\end{abstract}
\begin{IEEEkeywords}
Sleep Stage Classification, Transfer Learning, EEG, Domain Alignment
\end{IEEEkeywords}
\section{Introduction}
Sleep staging is crucial for assessing sleep quality and diagnosing disorders, typically involving the use of biometric signals to categorize sleep stages (N1, N2, N3, and REM). Sleep experts rely on polysomnography (PSG) data for judgments, including EEG and other physiological signals, but manual evaluation is costly and prone to errors \cite{b1, b3}. As a result, automated sleep staging methods have been extensively studied.

Deep learning has been widely applied to automatic sleep stage classification, showing great performance in improving the accuracy and efficiency of sleep studies\cite{b50,b17,zhou2024bit,jia2022hybrid,zhou2023interpretable}. However, in practical sleep staging, there are often situations where judgments need to be made directly on small groups of individuals, and specifically training a model for these data is not cost-effective. For instance, some sleep studies may involve only a small cohort of ten to twenty subjects, making it impractical to invest significant resources in developing a model tailored exclusively to such limited data \cite{b4,b5,b6,b7,zhou2023eeg}. Additionally, the process of training with large datasets is labor-intensive and resource-intensive, involving complex steps from data preprocessing to model training and validation.  Thus, reducing training efforts while maintaining  the model's accuracy on new samples is important.

Transfer learning offers a solution, where a model is initially trained on a large source domain dataset and then be finetuned with a small target domain dataset \cite{b13,liuvbh}. However, when knowledge from the source domain negatively impacts target domain learning, it results in negative transfer, reducing model performance \cite{b8,b11}.

Obtaining sufficient labeled data in the target domain is challenging due to the high costs involved, even though labeled data is generally available in the source and target domain \cite{b12}. Furthermore, even if enough target domain data is collected, domain shifts often occur due to variations in EEG signals caused by factors such as physiology, age, and health \cite{b14}, rendering models trained on the source domain less applicable. To mitigate this issue, we adopt an unsupervised approach that uses only labeled source data, carefully selecting source data that closely aligns with the target domain. This approach reduces domain mismatch, minimizes the risk of negative transfer, and enhances the applicability of knowledge transferred from the source domain to the target domain.

In this paper, we propose an domain aligning approach for sleep staging to maintain model performance on new data by mitigating negative transfer. Our Domain Aligner selects high-quality source data for transfer learning, making batch-based decisions to improve efficiency. We use specific metrics to identify similar source data and update the pre-trained model's parameters. Our contributions include:
\begin{itemize} \item We introduce a Domain Aligner designed to identify the most suitable source data, thereby helping the model mitigate the effects of negative transfer. \item Rather than generating new features and selecting data on a case-by-case basis, our approach directly utilizes the source domain data and processes it in batches, improving the model's training efficiency. \item Experimental results demonstrate that our Domain Aligner enhances the performance of the transfer learning model in sleep staging. Moreover, our model achieves state-of-the-art (SOTA) performance on sleep staging tasks. \end{itemize}
\begin{figure*}
\centering
\includegraphics[width=0.8\linewidth]{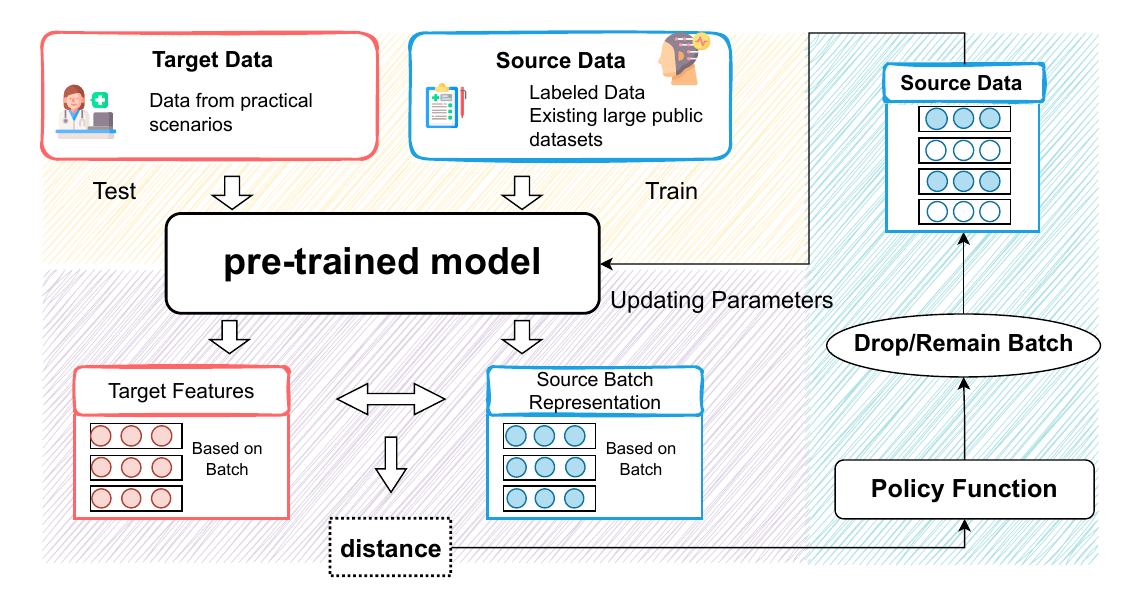} 
\caption{An overview of our SelectiveFinetuning, which composes of Pre-trained Model, Feature Extraction and Domain Aligner. It extracts features from both source and target domain data using a pre-trained model and calculates their similarity. Then, a policy function selects the source domain data that is most similar to the target domain. These selected data are then used to finetune the model's parameters, reducing the impact of negative transfer and ultimately improving the model's performance on the target domain.}
\label{fig:model-architecture}
\end{figure*}
\section{Related Work}
\subsection{Sleep Staging}

Currently, deep learning-based sleep staging networks typically consist of feature extraction and sequential signal classification. For example, DeepSleepNet uses CNN for feature extraction and BiLSTM for classification \cite{b4}; SeqSleepNet extracts time-frequency features from EEG signals using a hierarchical RNN \cite{b18}; XSleepNet jointly learns from raw signals and time-frequency images with a sequence-to-sequence model \cite{phan2021xsleepnet}.

While all methods have demonstrated high efficacy when provided with ample labeled training data, practical sleep staging often involves evaluating data from a limited number of subjects. This paucity of data is insufficient for training accurate models \cite{b32,b33}. Therefore, it is necessary to employ transfer learning to leverage the knowledge the model has acquired from large datasets \cite{b34}.
\subsection{Transfer Learning}
Transfer learning can be categorized into three types based on label-setting: inductive, transductive, and unsupervised transfer learning \cite{b38,b37,b19}. For sleep staging, several studies have explored training personalized deep learning models \cite{b20,b35}. 

However, due to the characteristics of EEG signals, it
usually leads to differences in data distributions between
the source and target domains, and these differences often result in poor performance of the model in the target domain \cite{b39,b40}. To address this, there are two common approaches: selecting data from the source domain to achieve a closer alignment of probability distributions between the source and target domains \cite{b22}, or defining a shared feature space to generate features that reduce the divergence between the two domains \cite{b23,b24}.

The method proposed in this paper follows a fully transductive transfer framework, where no labels from the target domain are used. We select data from the source domain that aligns well with the target domain using features from a common feature space. This approach effectively mitigates negative transfer and can replace the need to collect large amounts of new data by finetuning with source domain data.

\section{Method}
The architecture of our algorithm framework is illustrated in Fig. \ref{fig:model-architecture}. Our framework is designed with the primary goal of addressing the issue of negative transfer in transfer learning.  We summarize three key ideas:
\begin{enumerate}
    \item Employ a pre-trained model that effectively extracts EEG signal features to provide a strong foundation for transfer learning.
    \item Design a feature extraction method that maps both source and target domain data into a common feature space.
    \item Develop a domain aligning mechanism that specifically aims to minimize negative transfer by carefully selecting source domain data.
\end{enumerate}

\subsection{Pre-trained Model}
Different sleep stages are associated with specific EEG signal frequencies, making it important to capture these varying bands for accurate classification. To achieve this, we apply a Multi-Resolution CNN model \cite{b31}. This model includes two convolutional branches: a wide kernel branch for lower frequencies like the delta ($\delta$) band and a narrow kernel branch for higher frequencies such as the alpha ($\alpha$) and theta ($\theta$) bands. This part is indicated in the yellow part of Fig. \ref{fig:model-architecture}.

The model extracts features using two convolutional branches: one with a wide 400 kernel for lower frequencies, and another with a narrow 50 kernel for higher frequencies. The operations are defined as:
\[
F_{s1} = \text{Conv}(X_s, K=400)
\]
\[
F_{s2} = \text{Conv}(X_s, K=50)
\]

This structure enables the model to capture a broad range of EEG signal frequencies, essential for robust performance in transfer learning tasks.

\subsection{Feature Extraction}
Because MRCNN effectively extracts distinct features for each sleep stage from EEG signals, we choose to utilize a pre-trained MRCNN as the feature extractor, mapping both source and target domain data into a common feature space, which is indicated in the purple part of Fig. \ref{fig:model-architecture}. 

The specific steps are as follows:

\begin{itemize}
    \item We extract the output features from a specific convolutional layer (denoted by \(\sigma\)) for both the source and target domain data, processing them in batches. This ensures uniformity in the feature extraction process across domains:
    \[
    F_s^{(l)} = \sigma\left(W_s^{(l)} F_s^{(l-1)} + b_s^{(l)}\right)
    \]
    \[
    F_t^{(l)} = \sigma\left(W_t^{(l)} F_t^{(l-1)} + b_t^{(l)}\right)
    \]
    
\item The final feature representations for the source and target domains are derived from the specific convolutional layer's outputs. By maintaining the same network and layer, we ensure a unified feature space:
\[
F_s = F_s^{(l)},\quad F_t = F_t^{(l)}
\]
\end{itemize}

This shared feature extraction strategy ensures consistency across domains, which can support subsequent alignment process.

\subsection{Domain Aligner}
To mitigate negative transfer, our Domain Aligner evaluates and selects source domain data that closely matches the target domain. This targeted selection ensures that the data used to selectively finetune the pre-trained model is most relevant.

The aligner (indicated in the green part of Fig. \ref{fig:model-architecture}) operates as follows:

\begin{itemize}
\item {Policy Function and Reward Calculation}:
\begin{itemize}
\item Let \(\phi\) denote the function that extracts features \(F_s\) and \(F_t\) from the source and target domains using the shared feature extraction method:

    \[
    F_s = \phi(X_s) ,\quad F_t = \phi(X_t)
    \]

\item The policy function calculates the similarity between the domains using Earth Mover's Distance (EMD). 
\[
\text{EMD} (F_s, F_t) = \min_{\{f_{ij}\}} \sum_{i=1}^{m} \sum_{j=1}^{n} f_{ij} \cdot d(f_s^i, f_t^j)
\]
subject to:
\[
\sum_{j=1}^{n} f_{ij} = w_s^i  ,\quad \sum_{i=1}^{m} f_{ij} = w_t^j, \quad f_{ij} \geq 0
\]

Here, \(d(f_s^i, f_t^j)\) denotes the distance between feature vectors \(f_s^i\) and \(f_t^j\), while \(f_{ij}\) denotes the optimal transport needed to map the source domain features \(F_s\) to the target domain features \(F_t\).

\item A reward \(R\) is assigned to each source domain sample, inversely proportional to the EMD, prioritizing samples more similar to the target domain:
\[
R = \frac{1}{\text{EMD}(F_s, F_t)}
\]
\end{itemize}

\item {Data Selection and Selectively Finetune}:
\begin{itemize}
    \item Source domain data with reward values \(R\) below a certain threshold \(\tau\) are selected:
    \[
    \text{Selected\_source\_data} = \{ F_s \mid R > \tau \}
    \]
    \item The selected samples are used to finetune the pre-trained model.
    \end{itemize}
\end{itemize}

\section{Experiments}
We perform an evaluation of our algorithm method in comparison with baseline models on the SleepEDF and SHHS datasets.
\subsection{Dataset and  Preprocessing}
The SleepEDF dataset\cite{b7} contains 153 sleep recordings from 78 healthy individuals aged 25 to 101 years, with most subjects having two full-day polysomnography (PSG) recordings. These are segmented into 30-second epochs, and labeled
according to sleep stages, following AASM guidelines, where
stages N3 and N4 are merged and Movement/Unknown stages
are excluded, resulting in the final set \{W, N1, N2, N3, REM\}.
EEG are taken from the FPZ-CZ channel. Similarly,
the SHHS dataset \cite{b25} includes PSG recordings from 329
participants, selected based on their Apnea-Hypopnea Index
(AHI), using the C4-A1 EEG channel, with data processed
similarly to SleepEDF.


For the experimental evaluation, a cross-dataset transfer learning setup is used. The model is trained on one dataset as the source domain and evaluated on the other as the target domain. To account for different sampling frequencies across channels, resampling is applied to standardize all data to 100 Hz.
\subsection{Baseline Methods}
We compare our method with the following six baseline methods:

No Adapt: A baseline where the model is pre-trained on source data and then tested on target domain data without any adaptation.





MSTGCN:  Combines multi-view spatiotemporal information to enhance the generalization in sleep stage classification\cite{b26}.

SEN-DAL: Utilizes multi-modal signals and domain adversarial learning to improve accuracy and generalization in sleep staging\cite{b27}.

TENT:Achieves test-time adaptation through entropy minimization, improving model performance in new environments\cite{b28}.

EATA: Dynamically adjusts the model at test time through closed-loop inference to enhance performance\cite{b29}.

SAR: Adapts to changes in dynamic environments with a stable test-time adaptation method\cite{b30}.

MUDA: The method uses domain-specific and domain-invariant branches with an adaptive mixing strategy to address domain shifts\cite{b41}.

TTACS: An OTTA framework with a teacher-student network to learn domain-invariant features \cite{b42}.
\subsection{Experiment Results}
\begin{table}[h]
\centering
\resizebox{0.7\linewidth}{!}{
\begin{tabular}{lcc}
\toprule
\textbf{Method} & \textbf{Accuracy} & \textbf{F1-score} \\ \midrule
SleepEDF-20     & {82.9}            & {78.5}     \\ 
SleepEDF-78     & 81.6              & 78.2              \\ 
SHHS            & 79.4              & 75.8              \\ \bottomrule
\end{tabular}
}
\vspace{0.2cm}
\caption{\textnormal{The results of 10-fold cross-validation using the MRCNN model, comparing the accuracy and F1-score metrics across three different datasets (SleepEDF-20, SleepEDF-78, and SHHS).}}
\label{tab:pretrain_mrcnn}
\end{table}
\begin{table}[h]
    \centering
    \renewcommand{\arraystretch}{1.2} 
    \resizebox{1.0\linewidth}{!}{
    \Large 
    \begin{threeparttable}
    \begin{tabular}{lcccc}
        \toprule
        \multirow{2}{*}{\parbox[c]{2cm}{\centering \textbf{Method}}} & \multicolumn{2}{c}{\textbf{SHHS $\rightarrow$ EDF78}} & \multicolumn{2}{c}{\textbf{SHHS $\rightarrow$ EDF20}} \\
        \cline{2-5}
         & \textbf{Accuracy} & \textbf{F1-score}            
         & \textbf{Accuracy} & \textbf{F1-score}           \\ 
        \midrule
        No Adapt Strategy    & 65.2              & 61.8                        & 65.8              & 63.1                        \\ 
        MSTGCN               & 67.6              & 62.3                        & 68.4              & 66.4                        \\ 
        SEN-DAL              & 65.3              & 56.4                        & 68.4              & 66.4                        \\ 
        TENT                 & 68.7              & 65.4                        & 69.3              & 68.3                        \\ 
        EATA                 & 67.3              & 66.9                        & 67.5              & 66.4                        \\ 
        SAR                  & 73.1              & 71.8                        & 74.1              & 71.9                        \\ 
        MUDA & 72.8              & 71.4                        & 73.4              & 72.2                        \\ 
        TTACS & 73.2              & 71.0                        & 73.8              & 71.6                        \\ 
        Ours                 & \textbf{75.2}     & \textbf{73.7}               & \textbf{76.7}     & \textbf{74.9}               \\ 
    \bottomrule
    \end{tabular}
   \end{threeparttable}  
   }
   \vspace{0.2cm}
    \caption{\textnormal{Performance of our method between the SleepEDF and SHHS datasets. The notation “A→B” signifies that the transfer learning method was applied to transfer knowledge from dataset A to dataset B.}}
    \label{tab:mutual_transfer}
    
\end{table}
Table \ref{tab:pretrain_mrcnn} demonstrates MRCNN's effectiveness, with strong accuracy and F1-scores across all datasets, particularly on SleepEDF-20. However, despite this strong performance, when MRCNN is transferred to a different dataset, its effectiveness diminishes, leading to a decline in performance.
These baseline algorithms, shown in Table \ref{tab:mutual_transfer}, can be classified into two categories: enhancing generalization in sleep stage classification and focusing on test-time adaptation.

MSTGCN and SEN-DAL improve generalization but face complexity issues, while TENT, EATA, and SAR enhance test-time adaptation with risks of forgetting and resource demands. MUDA struggles with domain shifts, and TTACS relies on batch-level data, making both less robust against imbalanced class distributions in sleep staging.

\begin{figure}[h]
    \centering
    \includegraphics[width=1.0\linewidth]{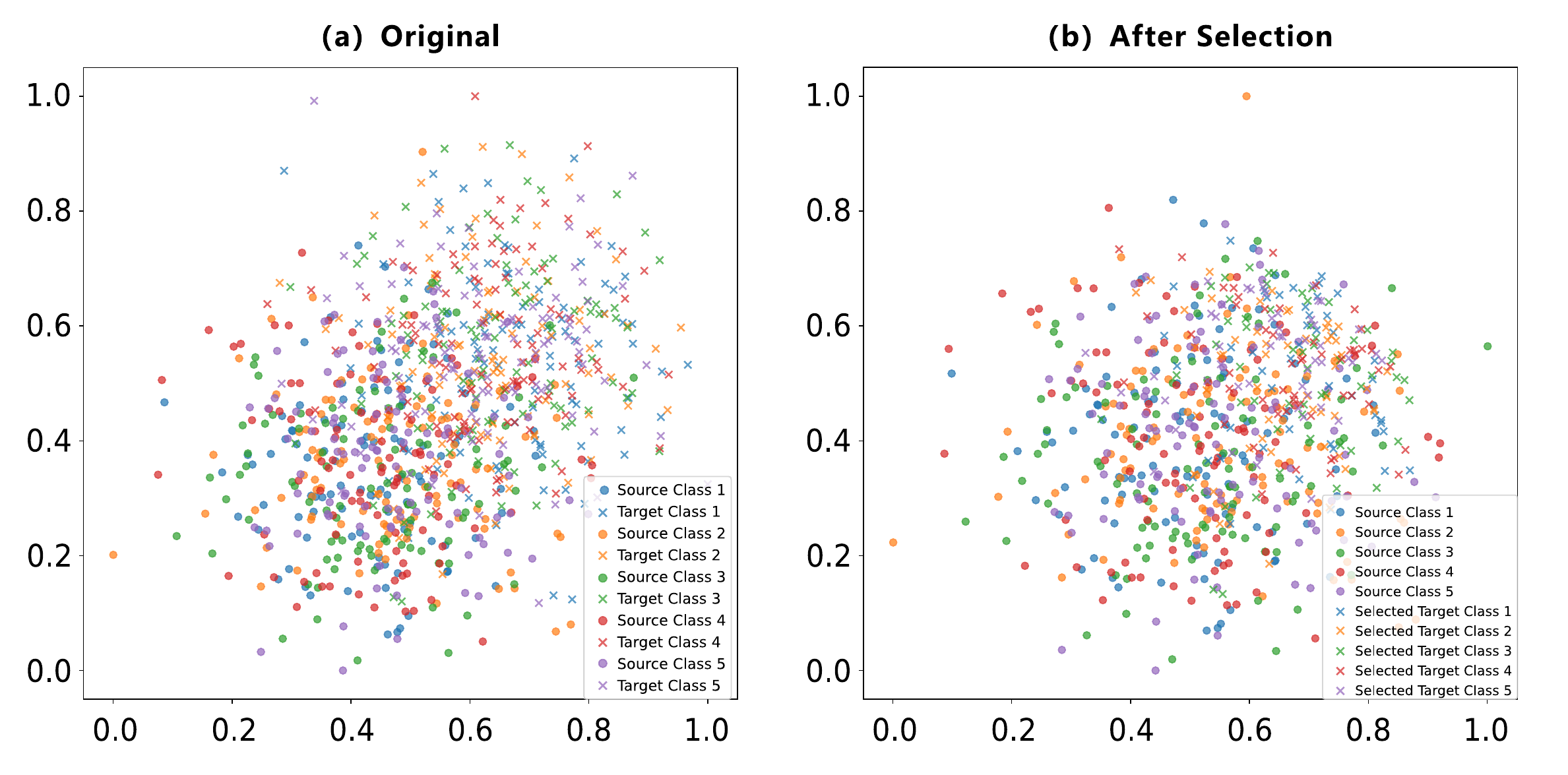} 
    \caption{\textnormal{This figure shows the effect of aligning data between the source and target domains in the presence of domain shifts. The left plot displays mixed data distributions, while the right plot shows a more aligned distribution after selecting source data similar to the target domain.} }
    \label{fig:sim-plot} 
\end{figure}

Among the compared methods, MSTGCN and TENT show moderate improvements, with TENT benefiting from entropy minimization. SAR demonstrates strong stability, achieving high F1-scores due to sharpness-aware optimization, while MUDA performs better overall but falls short of the highest accuracy and F1-scores achieved by our method. The key advantage of our method is its ability to minimize domain shifts by selectively finetuning on relevant source data (as shown in Fig. \ref{fig:sim-plot}), which enhances the model's adaptation to new datasets and reduces the risk of negative transfer. Consequently, our method outperforms others, achieving the highest accuracy and F1-scores in both the SHHS → 78 and SHHS → 20 tasks.
\section{Conclusion}
In this paper, we propose a method to mitigate negative transfer in sleep stage classification by strategically selecting source domain data that closely resembles the target domain. Our approach begins with a pre-trained MRCNN to extract EEG features and map them into a common feature space. We then employ a domain alignment mechanism that uses EMD to assess the distributional differences between source and target domain data. Based on this evaluation, source domain data that best matches the target domain is selected and used to finetune the pre-trained model. Our Domain Aligner allows the model to adapt more effectively to the target domain, reducing the impact of domain shifts and improving performance, particularly in scenarios with limited or small sample data. By addressing domain shifts in this way, our approach enhances the robustness, accuracy, and adaptability of sleep stage classification, thus making it a valuable tool for practical applications.
\bibliographystyle{IEEEtran}
\bibliography{ref}
\end{document}